\documentclass[pra,prl,twocolumn,showpacs,floatfix,10pt]{revtex4-1}
\usepackage{graphicx}
\usepackage{dcolumn}
\usepackage{bm}

\usepackage{float}
\usepackage{subfigure}
\usepackage{amsmath}
\usepackage{latexsym,amssymb,graphicx}
\usepackage[mathscr]{eucal}
\usepackage[bookmarks=true,
   colorlinks=true,
   linkcolor=blue,
   urlcolor=magenta,
   citecolor=magenta,
   bookmarks=true,
   hyperindex=true
]{hyperref}

\begin{document}
\allowdisplaybreaks[2]

\title{The thermodynamics and phase transition of  a rainbow black hole}

\author{Zhong-Wen Feng\textsuperscript{1}}
\altaffiliation{Email: zwfengphy@163.com}
\author{De-Ling Tang\textsuperscript{2}}
\altaffiliation{Email: cwnutangdl@163.com}
\author{Dan-Dan Feng\textsuperscript{2}}
\altaffiliation{Email: ncddfeng@163.com}
\author{Shu-Zheng Yang\textsuperscript{1}}
\altaffiliation{Email: szyangphys@126.com}
\vskip 0.5cm
\affiliation{1 College of Physics and Space Science, China West Normal University, Nanchong 637009, China\\
2 School of Science and Engineering, Sichuan Minzu College, Kangding 626001, China}

\date{\today}

\begin{abstract}
In this work, we construct a new kind of rainbow functions, which has generalized uncertainty principle parameter. Then, we investigate modif\/ied thermodynamic quantities and phase transition of rainbow Schwarzschild black hole by employing this new kind of rainbow functions. Our results demonstrate that the rainbow gravity and generalized uncertainty principle have a great ef\/fect on the picture of Hawking radiation. They prevent black holes from total evaporation and cause a remnant. In addition, after analyzing the modif\/ied local thermodynamic quantities, we f\/ind that the ef\/fect of rainbow gravity and the generalized uncertainty principle lead to one f\/irst-order phase transition, two second-order phase transitions, and two Hawking-Page-type phase transitions in the thermodynamic system of rainbow Schwarzschild black hole.
\end{abstract}

\maketitle

\section{Introduction}	
\label{Int}
In the theoretical physics, one of the biggest challenges is to combine the quantum theory with gravity theory. In order to achieve this purpose, various quantum gravity theories, such as the string theory, loop quantum gravity, M-theory, and non-commutative geometry have been proposed \cite{ch1,ch1+,ch2+,ch2,ch3}. According to those theories, it is found that the existence of a minimum measurable length, which is expected to be of the order of the Planck scale. However, under the linear Lorentz transformations, one f\/inds that the Planck length is variable, which is conf\/lict with the previous conclusions. To resolve this contradiction, the standard energy-momentum dispersion relation needs to be modif\/ied \cite{ch5,ch6,ch7}. Nowadays, the modif\/ications of standard dispersion relation are called as modif\/ied dispersion relation (MDR) \cite{ch8,ch9,ch10}. Interestingly, it should be noted that the MDR can modify the special relativity. In Ref.~\cite{ch11}, this generalization of special relativity is called as double special relativity (DSR), which has two invariants: the velocity of light $c$, and the Planck length $\ell_p$ (or Planck energy $E_p$).  However, DSR still two f\/laws in the theory: one is the DSR can only be used in f\/lat spacetime, the other is the dual position space in this theroy def\/inition suf\/fers a nonlinearity of the Lorentz transformation. Therefore,  Magueijo and Smolin generalized the DSR into curved spacetime and proposed the rainbow gravity (RG) \cite{ch12}. In the RG, the energy of a test particle af\/fects the geometry. In other word, the spacetime cannot be described by a single metric, but a family of metrics (rainbow of metrics) parameterized by the ratio ${E \mathord{\left/ {\vphantom {E {E_p }}} \right. \kern-\nulldelimiterspace} {E_p }}$. According to the RG, the deformed energy-momentum relation is given by
\begin{align}
\label{eq1}
E^2 f^2 \left( {{E \mathord{\left/ {\vphantom {E {E_p }}} \right. \kern-\nulldelimiterspace} {E_p }}} \right) - p^2 g^2 \left( {{E \mathord{\left/
 {\vphantom {E {E_p }}} \right. \kern-\nulldelimiterspace} {E_p }}} \right) =m^2,
\end{align}
where  $E_p$, $p$ and $m$ represent the Planck energy,  the momentum and mass of a test particle, respectively. The correction terms  $f\left( {{E \mathord{\left/{\vphantom {E {E_p }}} \right. \kern-\nulldelimiterspace} {E_p }}} \right) $ and $g\left( {{E \mathord{\left/ {\vphantom {E {E_p }}} \right. \kern-\nulldelimiterspace} {E_p }}} \right)$ are the rainbow functions, which satisfy the limits $\mathop {\lim }\limits_{{E \mathord{\left/ {\vphantom {E {{E_p} \to 0}}} \right. \kern-\nulldelimiterspace} {{E_p} \to 0}}} f\left( {{E \mathord{\left/
 {\vphantom {E {{E_p}}}} \right. \kern-\nulldelimiterspace} {{E_p}}}} \right) = 1$ and $\mathop {\lim }\limits_{{E \mathord{\left/ {\vphantom {E {E_p }}} \right. \kern-\nulldelimiterspace} {E_p }} \to 0} g\left( {{E \mathord{\left/  {\vphantom {E {E_p }}} \right. \kern-\nulldelimiterspace} {E_p }}} \right) = 1$. Those relations show that the RG is constraint to reproduce the standard dispersion relation in low energy limit. According to energy dependent modif\/ication to the dispersion relation, metric in RG is given by $g(E)=\eta^{a b} e_{a}(E) \otimes e_{b}(E)$ with the energy dependence of the frame f\/ields  $e_{0}\left(E / E_{P l}\right)=\left(1 / f\left(E / E_{P l}\right)\right) \tilde{e}_{0}$ and $e_{i}\left(E / E_{p}\right)=\left(1 / f\left(E / E_{p}\right)\right) \tilde{e}_{i}$.

In recent years, the implications of RG have been investigated in many contexts, such as cosmology, astrophysics, and black hole physics \cite{ch13,ch14,ch15,che1,che2}. In particular, the RG has a very signif\/icant ef\/fect on the thermodynamics of black holes. In Refs.~\cite{ch16,ch17,ch18,ch19,chf0,chf0+,chf1,chf1+,chf1++,chf2,chf3,chf3+,chf4,chf5,chf6,chf7,chf8}, the authors studied the modif\/ied thermodynamics of black holes by using the RG. These results demonstrated that the modif\/ications are departed from the classical cases when the size of black holes approach the Planck scale. It indicates that the RG can stop the Hawking radiation in the last stages of black holes' evolution and naturally leads to remnants. Therefore, the RG can used to address the information loss and naked singularity problems of black holes. Recently, according to the rainbow functions that were proposed by Magueijo and Smolin, we discussed the thermodynamics and phase transition of a spherically symmetric black hole. Our calculations showed that the RG changes the picture of black hole's thermodynamic phase transition~\cite{cha1}. In addition, we found three Hawking-Page-type phase transitions in the framework of the RG.

On the other hand, the minimum measurable length can also change the conventional Heisenberg uncertainty principle (HUP) to the so-called generalized uncertainty principle (GUP). Since the GUP can be applied to many physical systems, it attracted people's attention now \cite{cha2,cha3,cha4,cha5,cha6,cha7,cha8,cha9}. In Ref.~\cite{cha2}, Adler, Chen and Santiago pointed out that uncertainties for position and momentum of the GUP (namely, $\Delta x$ and $\Delta p$) can be considered as the temperature and radius of the black holes. With this heuristic theory, they studied the GUP corrected Hawking temperature of Schwarzschild (SC) black hole and showed that the modif\/ied Hawking temperature of SC black hole is higher than that of conventional case. However, in our previous work, by using the GUP together with Hamilton-Jacobi equation, the GUP corrected thermodynamics of SC black hole are similar to the results obtained in the framework of RG. This indicates that there has some connection between the GUP and RG \cite{cha10}. From the above discussion, it is interesting to incorporate the GUP with RG to investigate the modif\/ied thermodynamics and phase transition of black holes, this may lead to some novel results. Following this line of research, we construct a new kind of rainbow functions, which contains the ef\/fect of GUP in this paper. Then, by using this new kind of rainbow functions, the modif\/ied thermodynamics, thermodynamic criticality and the phase transition of rainbow SC black hole will be analyzed.

The remainders of this paper are outlined as follows. In Section~\ref{II}, we derive a MDR via the GUP that proposed by Kempf, Mangano, and Mann. Then, comparing this MDR with the general form of RG, a new kind of rainbow functions, which contains the GUP parameter is obtained. In Section~\ref{III}, according to the new kind of rainbow functions, we calculate the modif\/ied thermodynamics of SC black hole in the context of RG. In Section~\ref{V}, we study the modif\/ied local thermodynamic quantities of rainbow SC black hole, and discuss the phase transition between the hot space and the states of black holes. Finally, the discussion is given in Section~\ref{Dis}.

Throughout the paper, the natural units of $k_B = c = 1$ are used

\section{The rainbow functions with ef\/fect of GUP}
\label{II}
To begin with, we need to recapitulate the way of getting a MDR, which inf\/luenced by the GUP. In Ref.~\cite{cha11}, according to the GUP that proposed Kempf, Mangano, and Mann, we derive a kind of MDR, and used it to show that the GUP parameters can be constrained by the gravitational wave event GW150914 and GW151226. The expression of the GUP is given by
\begin{equation}
\label{eq2}
\Delta x\Delta p \ge \frac{\hbar }{2}\left[ {1 + \beta \left( {\Delta p} \right)^2 } \right],
\end{equation}
where $\Delta x$  and $\Delta p$  are the uncertainties for position and momentum, respectively. $\beta  = \beta _0 \ell _p^2 {\rm{ }} = {{\beta _0 } \mathord{\left/ {\vphantom {{\beta _0 } {M_p^2 }}} \right. \kern-\nulldelimiterspace} {M_p^2 }} = {{\beta _0 } \mathord{\left/ {\vphantom {{\beta _0 } {E_p^2 }}} \right. \kern-\nulldelimiterspace} {E_p^2 }}$ with the GUP parameters $\beta _0$, and $\ell _p$, $M _p$ ,and $E_p$ are represent the Planck length, Planck mass, and Planck energy, respectively. \cite{cha12}. Based on Eq.~(\ref{eq2}), one can easy obtain a nonzero minimal uncertainty $\Delta x_{\min }  \approx \ell _p \sqrt {\beta _0 }$. Furthermore, the operators for the modif\/ied position operator $x_i$ and modif\/ied momentum $p_i$ in Eq.~(\ref{eq2}) can be def\/ined as
\begin{equation}
\label{eq3}
x_i  = x_{0i} ; \quad {p_i} = {p_{0i}}\left( {1 + \beta {{\textbf{k}}^2}} \right),
\end{equation}
where  usual position operator $x_{0i}$ and usual momentum $p_{0i}$ are satisfying the canonical commutation relations $\left[ {x_{0i} ,p_{0j} } \right] = i\hbar \delta _{ij}$, ${\textbf{k}^2} = {\delta _{ij}}{k^{0i}}{k^{0j}}$, thus $\textbf{k} = \sqrt {{\delta _{ij}}{k^{0i}}{k^{0j}}}$. When considering a general gravitational background metric ${\rm{d}}s^2  = g_{ab} {\rm{d}}x^a {\rm{d}}x^b  = g_{00} {\rm{d}}t^2  + g_{ij} {\rm{d}}x^i {\rm{d}}x^j$ the modif\/ied square of the four-momentum can be expressed as \cite{cha13}
\begin{equation}
\label{eq4}
p_a p^a  = g_{ab} p^a p^b  =  g_{00} \left( {p^{00} } \right)^2  + g_{ij} p^{0i} p^{0j} \left( {1 + \beta \textbf{k}^2 } \right)^2.
\end{equation}
It is normally set that $g_{00}  =  - 1$, in which case the spacetime of MDR is f\/lat. Next, by ignoring the higher order term ${\cal O}\left( {\beta } \right)$, the equation above can be rewritten as
\begin{equation}
\label{eq5}
p_a p^a  =  - \left( {p^{00} } \right)^2  + \textbf{k}^2  + 2\beta \textbf{k}^2 \textbf{k}^2.
\end{equation}
In the right side of Eq.~(\ref{eq5}), the f\/irst two terms form the standard dispersion relation, that is $ - \left( {p^{00} } \right)^2  + \textbf{k}^2  = -m^2$. Thus, Eq.~(\ref{eq5}) can be rewritten as $p_a p^a  = -m^2 + 2\beta \textbf{k}^2 \textbf{k}^2$, and its time component of the momentum becomes
\begin{equation}
\label{eq6}
\left( {p^{00} } \right)^2  = m^2 + \textbf{k}^2 \left( {1 - 2\beta \textbf{k}^2 } \right).
\end{equation}
Using the def\/inition of a particle's energy $E = p^{00}$, the energy of a particle can be expressed in terms of the three spatial momentum and mass as follows
\begin{equation}
\label{eq7}
E^2  - \textbf{k}^2 \left( {1 - 2\beta \textbf{k}^2 } \right) = m^2.
\end{equation}
Now, for simplify the next research, we assume the mass of the test particle is zero, so that the momentum  in Eq.~(\ref{eq7}) satisfying the standard massless energy-momentum dispersion relation, that is $E^2=\textbf{k}^2$. Furthermore, in order to obtain a MDR that consistent with Eq. ~(\ref{eq1}), we need retained the form of  $\textbf{k}^2$  outside parentheses and replaced  $\textbf{k}^2$ in parentheses with the standard massless energy-momentum dispersion.  F\/inally, the Eq.~(\ref{eq7}) can be rewritten as \cite{chf4}
\begin{equation}
\label{eq8}
E^2  - p^2 \left( {1 - 2{{\beta _0 E^2 } \mathord{\left/  {\vphantom {{\beta _0 E^2 } {E_p^2 }}} \right. \kern-\nulldelimiterspace} {E_p^2 }}} \right) = 0.
\end{equation}
Next, comparing Eq.~(\ref{eq8}) with Eq.~(\ref{eq1}),  a new kind of rainbow functions is obtained as follows:
\begin{equation}
\label{eq9}
{f\left( {{E \mathord{\left/ {\vphantom {E {E_p }}} \right. \kern-\nulldelimiterspace} {E_p }}} \right) = 1,}   \quad  g\left( {{E \mathord{\left/
 {\vphantom {E {{E_p}}}} \right. \kern-\nulldelimiterspace} {{E_p}}}} \right) = \sqrt {1 - {{2{\beta _0}{E^2}} \mathord{\left/ {\vphantom {{2{\beta _0}{E^2}} {E_p^2}}} \right.
 \kern-\nulldelimiterspace} {E_p^2}}} .
\end{equation}
Now, by employing the GUP~(\ref{eq2}) and a general gravitational background metric, we construct a new kind of rainbow functions~(\ref{eq9}). It is obviously that Eq.~(\ref{eq9}) contains the GUP parameter $\beta _0$ and the ratio ${E \mathord{\left/{\vphantom {E {E_p }}} \right.\kern-\nulldelimiterspace} {E_p }}$, which indicates that it inf\/luenced by the ef\/fect of GUP and RG. In addition, Eq.~(\ref{eq9}) satisfying the conditions of $\mathop {\lim }\limits_{{E \mathord{\left/  {\vphantom {E {E_p }}} \right. \kern-\nulldelimiterspace} {E_p }} \to 0} f\left( {{E \mathord{\left/ {\vphantom {E {E_p }}} \right. \kern-\nulldelimiterspace} {E_p }}} \right) = 1$  and $\mathop {\lim }\limits_{{E \mathord{\left/ {\vphantom {E {E_p }}} \right. \kern-\nulldelimiterspace} {E_p }} \to 0} g\left( {{E \mathord{\left/ {\vphantom {E {E_p }}} \right. \kern-\nulldelimiterspace} {E_p }}} \right) = 1$ at low energies. In the subsequent discussions, by incorporating the rainbow functions~(\ref{eq9}) with a specif\/ic line element, we investigate the modif\/ied thermodynamics of a rainbow black hole.

\section{The modif\/ied thermodynamics of Schwarzschild black hole in the frame work of RG}
\label{III}
In this section, by utilizing the rainbow functions~(\ref{eq9}) with the def\/inition of surface gravity and f\/irst law of thermodynamics,  the modif\/ied thermodynamics of rainbow SC black hole are calculated. In Ref.~\cite{ch12}, the line element of rainbow SC black hole is given by
\begin{align}
\label{eq10}
 {\rm{d}}s^2  & =  A\left( {r,E,{E_p}} \right){\rm{d}}t^2  + B\left( {r,E,{E_p}} \right){\rm{d}}r^2  + C\left( {r,E,{E_p}} \right){\rm{d}}\Omega ^2
\nonumber \\
& =  - \frac{{1 - \left( {{{2GM} \mathord{\left/ {\vphantom {{2GM} r}} \right. \kern-\nulldelimiterspace} r}} \right)}}{{{f^2}\left( {{E \mathord{\left/  {\vphantom {E {{E_p}}}} \right. \kern-\nulldelimiterspace} {{E_p}}}} \right)}}{\rm{d}}{t^2} + \frac{{{{\left[ {1 - \left( {{{2GM} \mathord{\left/  {\vphantom {{2GM} r}} \right.  \kern-\nulldelimiterspace} r}} \right)} \right]}^ - }^1}}{{{g^2}\left( {{E \mathord{\left/  {\vphantom {E {{E_p}}}} \right.  \kern-\nulldelimiterspace} {{E_p}}}} \right)}}{\rm{d}}{r^2} + \frac{{{r^2}}}{{{g^2}\left( {{E \mathord{\left/  {\vphantom {E {{E_p}}}} \right.  \kern-\nulldelimiterspace} {{E_p}}}} \right)}}{\rm{d}}{\Omega ^2},
\end{align}
where $A\left( {r,E,{E_p}} \right)$, $B\left( {r,E,{E_p}} \right)$ and $C\left( {r,E,{E_p}} \right)$ represent the geometric properties of the rainbow black hole, $\rm{d}\Omega ^2 $ denotes the line elements of 2-dimensional hypersurfaces, which can be expressed as $\rm {d}\theta ^2  + \sin ^2 \theta \rm {d} \phi ^2$ in this work. Obviously, metric~(\ref{eq10}) has only one root of  $r_H  = 2GM$ satisfying $A\left( {r_H} \right) = 0$. For $f\left( {{E \mathord{\left/ {\vphantom {E {E_p }}} \right. \kern-\nulldelimiterspace} {E_p }}} \right) = g\left( {{E \mathord{\left/ {\vphantom {E {E_p }}} \right. \kern-\nulldelimiterspace} {E_p }}} \right) = 1$, Eq.~(\ref{eq10}) reduces to the original metric of SC black hole. According to the line element of rainbow SC black hole, the surface gravity on the event horizon is given by
\begin{equation}
\label{eq11}
\kappa  =  - \frac{1}{2}\mathop {\lim }\limits_{r \to r_H } \sqrt { - \frac{{g^{11} }}{{g^{00} }}} \frac{{\partial _r \left( {g^{00} } \right)}}{{g^{00} }} = \frac{{g\left( {{E \mathord{\left/ {\vphantom {E {E_p }}} \right. \kern-\nulldelimiterspace} {E_p }}} \right)}}{{f\left( {{E \mathord{\left/ {\vphantom {E {E_p }}} \right. \kern-\nulldelimiterspace} {E_p }}} \right)}}\frac{1}{{4 G M}}.
\end{equation}
It is clear that the original surface gravity $\kappa _0  = {1 \mathord{\left/ {\vphantom {1 {4GM}}} \right. \kern-\nulldelimiterspace} {4GM}}$ is modif\/ied by the rainbow functions. Using Eq.~(\ref{eq11}), one can easily obtain the deformed Hawking temperature of SC black hole
\begin{equation}
\label{eq12}
T_H^{{\rm{modif\/ied}}}   =  \frac{\kappa }{{2\pi }} = {\rm{ }}\frac{{g\left( {{E \mathord{\left/  {\vphantom {E {E_p }}} \right.  \kern-\nulldelimiterspace} {E_p }}} \right)}}{{f\left( {{E \mathord{\left/  {\vphantom {E {E_p }}} \right.  \kern-\nulldelimiterspace} {E_p }}} \right)}} T_H  = \sqrt {1 - 2\beta _0 \frac{{E^2 }}{{E_p^2 }}} \frac{1}{{8\pi GM}} ,
\end{equation}
where $ T_H  = {1 \mathord{\left/ {\vphantom {1 {8\pi GM}}} \right. \kern-\nulldelimiterspace} {8\pi GM}}$ is the original Hawking temperature of SC black hole. One can see that the modif\/ied Hawking temperature is depended on the GUP parameter $\beta_0$, the ratio  ${E \mathord{\left/ {\vphantom {E {E_p }}} \right.  \kern-\nulldelimiterspace} {E_p }}$ and the mass of SC black hole, which means that the GUP and RG have a very signif\/icant ef\/fect on the thermodynamics of black hole. When  $\beta_0\rightarrow0$ or  ${E \mathord{\left/ {\vphantom {E {E_p }}} \right. \kern-\nulldelimiterspace} {E_p }} \to 0$, the original Hawking temperature is recovered.

Here, we have two schemes to eliminate the energy $\omega$ dependence of the probe in Eq.~(\ref{eq12}). In scheme I, the HUP $\Delta x \ge {1 \mathord{\left/ {\vphantom {1 {\Delta p}}} \right. \kern-\nulldelimiterspace} {\Delta p}}$ still holds in the RG. Therefore, the HUP can be translated into a lower bound on the energy of a particle emitted in Hawking radiation, that is, $E \ge {1 \mathord{\left/ {\vphantom {1 {\Delta x}}} \right.  \kern-\nulldelimiterspace} {\Delta x}}$ \cite{cha2}. Then, in the vicinity of the black hole horizon, the uncertainty in the position can be taken to be the radius of the SC black hole  $\Delta x \approx r_H= {1 \mathord{\left/ {\vphantom {1 {2GM}}} \right. \kern-\nulldelimiterspace} {2GM}}$. Thus, one has the following relation
\begin{equation}
\label{eq13}
E \ge \frac{1}{{r_H }} = \frac{1}{{2GM}}.
\end{equation}
Substituting Eq.~(\ref{eq13}) into Eq.~(\ref{eq12}), the modif\/ied Hawking temperature becomes
\begin{equation}
\label{eq14}
T_H^{{\rm{modif\/ied}}}  = \sqrt {1 - \frac{{\beta _0 }}{{2G^2 M^2 E_p^2 }}} T_H  =  \sqrt {1 - \beta _0 \frac{{M_p^2 }}{{2M^2 }}} \frac{1}{{8\pi GM}},
\end{equation}
where in above equation the relation  $E_p^{-2}  = M_p^{-2}  = G$ is used. It should be noted that the expression in the square root can not be less than zero, namely, $1 - {{\beta _0 M_p^2 } \mathord{\left/ {\vphantom {{\beta _0 M_p^2 } {2M^2 }}} \right. \kern-\nulldelimiterspace} {2M^2 }} \ge 0$. Therefore, one can obtain a minimum mass of rainbow SC black hole, namely, ${M_{\min }} = {M_p}\sqrt {{{{\beta _0}} \mathord{\left/ {\vphantom {{{\beta _0}} 2}} \right. \kern-\nulldelimiterspace} 2}}$. The minimum mass of rainbow SC black hole shows that the minimum mass of rainbow SC black hole is related to $\beta_0$ and  $M_p$, it  indicates that the rainbow SC black hole has a remnant, which is of the order of Planck scale and can be considered as a candidate of dark matter.

\begin{figure}[htbp]
\centering
\subfigure[Scheme I]{\includegraphics[height=6cm,width=7cm]{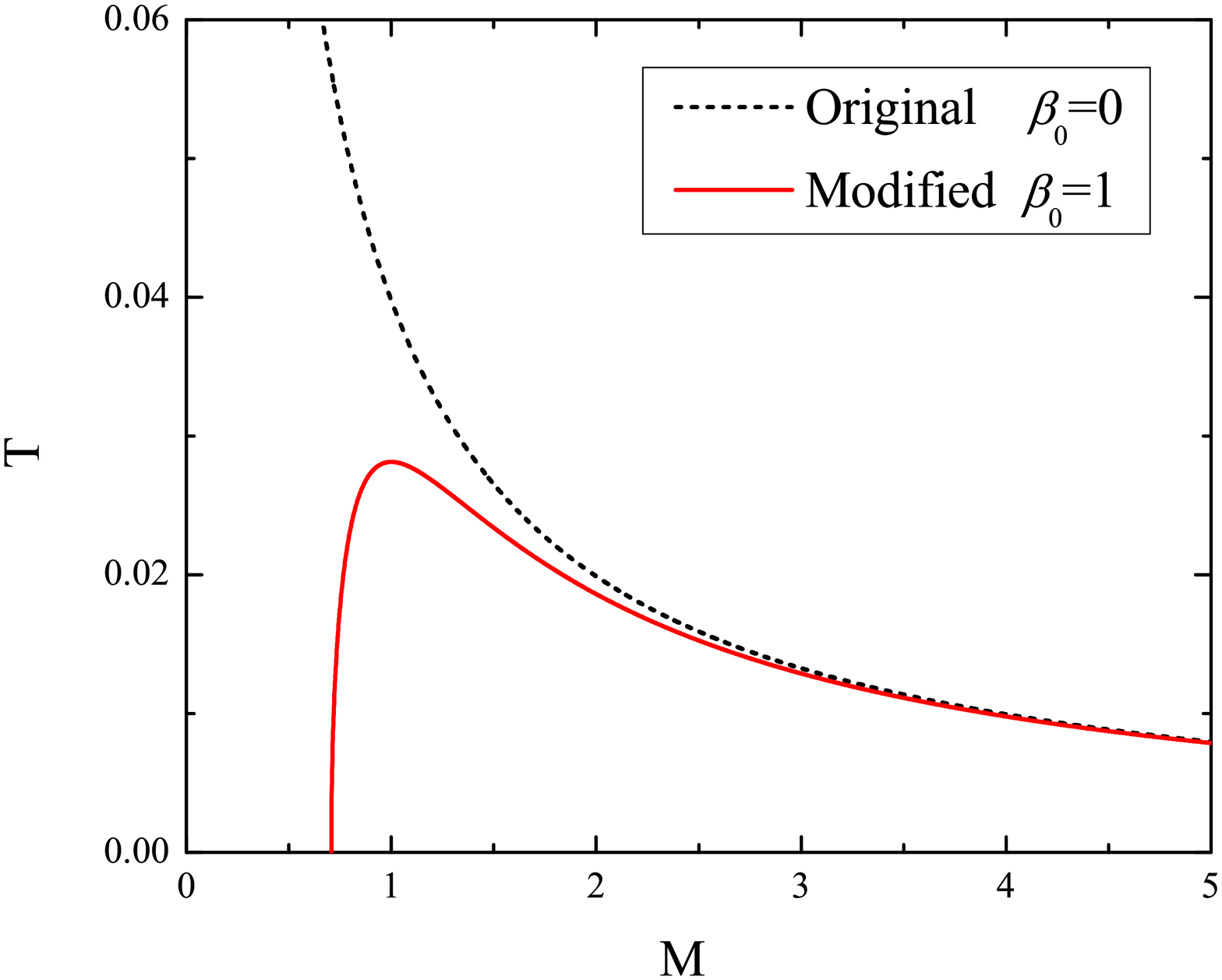}}
\subfigure[Scheme II]{\includegraphics[height=6cm,width=7cm]{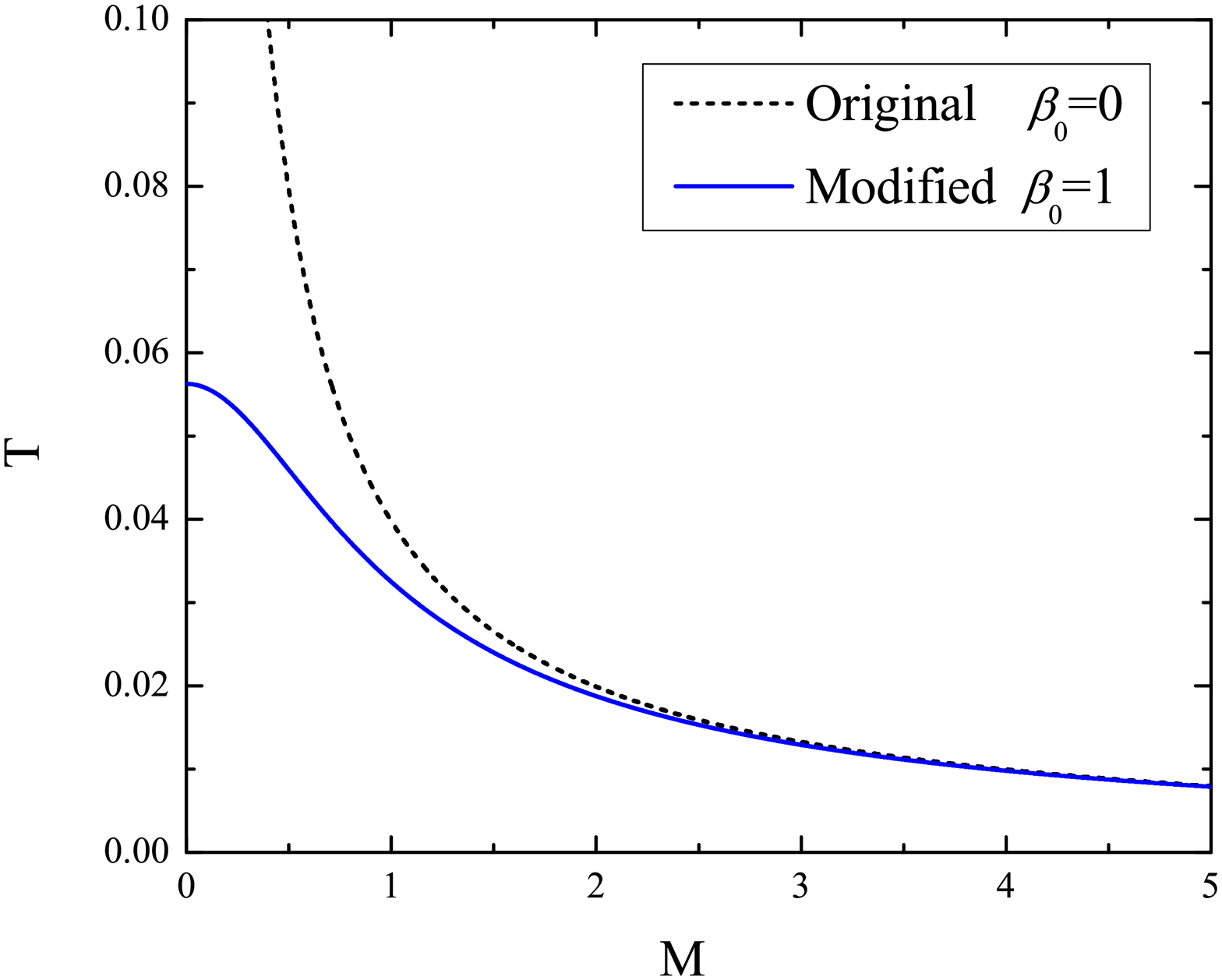}}
\caption{ The $T-M$ diagram of original SC black hole and rainbow SC black hole, where  $G$ and $M_p$  are equal to 1.}
\label{fig1}
\end{figure}

In scheme II, one can substitute the relation $\Delta p \sim {1 \mathord{\left/ {\vphantom {1 {{r_H}}}} \right. \kern-\nulldelimiterspace} {{r_H}}}$ and MDR~(\ref{eq1}) into Eq.~(\ref{eq12}) \cite{chf0,chf0+,chf1,chf1+,chf1++,chf2,chf3,chf3+,chf4,chf5,chf6,chf7,chf8,chx2,chx3}, the modif\/ied Hawking temperature becomes
\begin{equation}
\label{eq15}
\tilde T_H^{{\rm{modif\/ied}}} = \sqrt {1 - \frac{{2{\beta _0}E_p^2}}{{2{\beta _0} + 4{G^2}{M^2}E_p^2}}} \frac{1}{{8\pi GM}}.
\end{equation}
Now, by using Eq.~(\ref{eq14}) and Eq.~(\ref{eq15}), the behaviour of the original Hawking temperature $T_H$ and modif\/ied Hawking temperature  of SC black hole is depicted in Fig.~\ref{fig1}.

In Fig.~\ref{fig1}(a), the black dotted curve represents the original Hawking temperature of SC black hole for $\beta_0=0$, whereas the red solid curve illustrates the modif\/ied Hawking temperature of SC black hole for $\beta_0=1$. This diagram shows that the original Hawking temperature tends to inf\/inity as  $M \to 0$. On the contrary, the modif\/ied Hawking temperature shows another image of the black hole thermodynamics evolution. At the beginning of the evolution, the behaviour of modif\/ied Hawking temperature and that of original Hawking temperature look very much alike, since the ef\/fect of quantum gravity is weak at large scale. However, as the evolution progresses, the mass of the black hole gradually decrease, the behaviour modif\/ied Hawking temperature deviated from the original case. It rises at f\/irst, and then decreases rapidly after it reaches a maximum value. Finally, the modif\/ied temperature approaches zero when the mass tends to its minimum value $M_{\min }$. However, the behaviour of modif\/ied Hawking temperature (blue solid curve)  in Fig.~\ref{fig1}(b) is dif\/ferent from that in Fig.~\ref{fig1}(a). In this case, $\tilde T_H^{{\rm{modif\/ied}}}$ becomes f\/inite when the mass of rainbow SC black hole goes to zero. Since the modif\/ied Hawking temperature $\tilde T_H^{{\rm{modif\/ied}}}$ and its corresponding thermodynamic phase transition have been studied in detail in Refs.~\cite{chx2,chx3,chx4},  we only focus on the Eq.~(\ref{eq14}) and its corresponding thermodynamic phase transition in upcoming contents.

\begin{figure}[htbp]
\centering
\includegraphics[width=.5\textwidth,origin=c,angle=0]{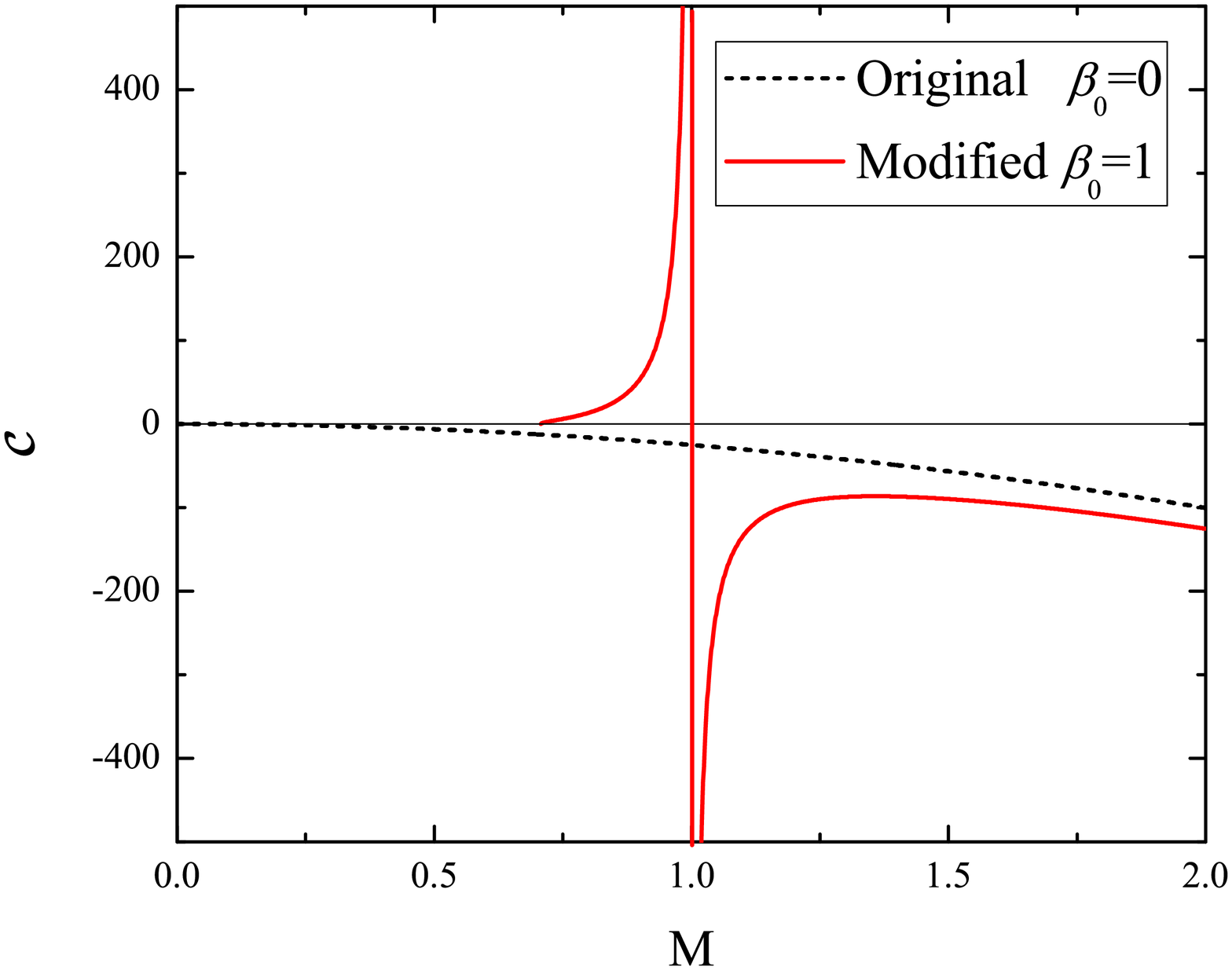}
\caption{\label{fig3} The $\mathcal{C}-M$ diagram of original SC black hole and rainbow SC black hole, where we set $G=M_p=1$.}
\end{figure}

Now, by using the area law \cite{ch36,ch21}, the modif\/ied entropy is
\begin{equation}
\label{eq16}
{S^{{\rm{modif\/ied}}}} = \frac{{2\pi G}}{{M\sqrt {2 - {{{\beta _0}M_p^2} \mathord{\left/ {\vphantom {{{\beta _0}M_p^2} {{M^2}}}} \right. \kern-\nulldelimiterspace} {{M^2}}}} }}\Xi ,
\end{equation}
where $\Xi  = \sqrt 2 M\left( {2{M^2} - {\beta _0}M_p^2} \right) + {\beta _0}M_p^2\sqrt {2{M^2} - {\beta _0}M_p^2}$ $ \ln \left( {2M + \sqrt {4{M^2} - 2{\beta _0}M_p^2} } \right)$.  When  $\beta_0  = 0$, the above equation becomes to the area law of the entropy of black hole $S = {A \mathord{\left/ {\vphantom {A 4}} \right. \kern-\nulldelimiterspace} 4} = 4\pi G M^2$ with the area of black hole $A$, this is consistent with the original entropy in Ref.~\cite{ch36}. Next, in order to determine whether the rainbow SC black hole exists a remnant, it is necessary to further analyze the specif\/ic heat of rainbow SC black. According to Ref.~\cite{cha10}, the def\/inition of specif\/ic heat is given by $\mathcal{C}= T_H \left( {{{\partial S} \mathord{\left/ {\vphantom {{\partial S} {\partial M}}} \right.  \kern-\nulldelimiterspace} {\partial M}}} \right)\left( {{{\partial T_H } \mathord{\left/ {\vphantom {{\partial T_H } {\partial M}}} \right. \kern-\nulldelimiterspace} {\partial M}}} \right)^{ - 1}$. Therefore, the modif\/ied specif\/ic heat is
\begin{equation}
\label{eq17}
\mathcal{C}^{{\rm{modif\/ied}}}  =  - \frac{{4\pi GM^4 }}{{M^2  - M_p^2 \beta _0 }}\sqrt {4 - \frac{{2M_p^2 \beta _0 }}{{M^2 }}} .
\end{equation}
From the equation above, it is found that $\mathcal{C}^{{\rm{modif\/ied}}}  = 0$ at  $M_{\min }  = M_p \sqrt {{{\beta _0 } \mathord{\left/
 {\vphantom {{\beta _0 } 2}} \right. \kern-\nulldelimiterspace} 2}}$, which is equal to Eq.~(\ref{eq15}). For $\beta_0=0$, Eq.~(\ref{eq17}) recovers the original case  ${\cal C}_0 =  - 8\pi M^2$. The associated $\mathcal{C}-M$ diagram is displayed in Fig.~\ref{fig3}.
One can see that the value of original specif\/ic heat (black dashed curve) is always negative, and it vanishes when $M \to 0$. Dif\/ferent from the original one, the modif\/ied specif\/ic heat (red solid curve) has a vertical asymptote at a certain point, where the temperature of rainbow SC black hole reaches its maximum value. It causes the value of specif\/ic heat to change from negative to positive, which implies a thermodynamic phase transition occurred at this certain location. The specif\/ic phase transition of rainbow SC black hole will be analyzed in the next section. In addition, it is clear that the modif\/ied specif\/ic heat vanishes when the mass of the reaches  $M_{\min }$. Since the black hole does not exchange their heat with the surrounding spacetime when the specif\/ic heat is equal to zero, we can determine the rainbow SC black hole has a remnant at $M_{\min }$. Therefore, the remnant mass is $M_{{\rm{res}}}  = M_{\min }$.

\section{Modif\/ied local thermodynamic quantities and phase transition of rainbow SC black hole}
\label{V}
\begin{figure}[htbp]
\centering
\includegraphics[width=.5\textwidth,origin=c,angle=0]{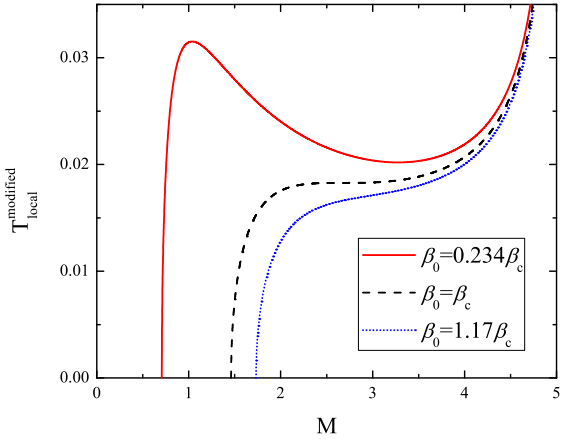}
\caption{\label{fig4} The $T_{{\rm{local}}}^{{\rm{modif\/ied}}}-M$ diagram of rainbow SC black hole for dif\/ferent $\beta_0$ with $\mathcal{R}=10$ and $G = M_p =1$.}
\end{figure}

It is well known that the black holes have rich phase structures and critical phenomena \cite{ch22,ch23,ch24,ch25,ch26,ch27,ch28,ch29,ch29+,ch30+}. The phase structures and critical phenomena of black holes have been a subject of fascination for more than forty years since they do not only of\/fer a deep insight into the quantum properties of gravity, but also allow for a deeper understanding of the thermal properties of conformal f\/ield theories. Therefore, in this section, for investigating the phase transition from the hot space to black hole, we need to further study the local thermodynamic quantities of the rainbow SC black hole.  In order to obtain the local thermodynamic quantities, the rainbow SC black hole needs to be located at the center of a spherical cavity of radius $\mathcal{R}$  \cite{ch37,ch38}. Hence, the expression of local temperature of rainbow SC black hole is given by
\begin{equation}
\label{eq18}
T_{{\rm{local}}}^{{\rm{modif\/ied}}}  = \sqrt {1 - \beta _0 \frac{{M_p^2 }}{{2M^2 }}} \frac{1}{{8\pi GM\sqrt {1 - \frac{{2GM}}{\mathcal{R}}} }}.
\end{equation}
As $\mathcal{R}\rightarrow \infty$, the above equation recovers the modif\/ied Hawking temperature. When considering the cavity of radius $\mathcal{R}$ as a invariable quantity, one can calculate the critical value of black hole's mass $M_{c}$, GUP parameter $\beta_{c}$, and local temperature $T_c$ by the following equations \cite{cha1}
\begin{equation}
\label{eq19}
{\left( {\frac{{\partial T_{{\rm{local}}}^{{\rm{modif\/ied}}} }}{{\partial M}}} \right)_\mathcal{R}  = 0,} \quad {\left( {\frac{{\partial ^2 T_{{\rm{local}}}^{{\rm{modif\/ied}}} }}{{\partial M^2 }}} \right)_\mathcal{R}  = 0}.
\end{equation}
Substituting Eq.~(\ref{eq18}) into Eq.~(\ref{eq19}), one can obtain the critical mass, critical GUP parameter, and critical local temperature as follows
\begin{equation}
\label{eq20}
{M_c  = \frac{{4\mathcal{R}}}{{15G}},} \quad {\beta _c  = \frac{{16\mathcal{R}^2 }}{{375G^2 M_p^2 }},} \quad {T_c  = \frac{{15}}{{32\pi \mathcal{R}}}\sqrt {\frac{3}{2}}.}
\end{equation}
For the convenience of discussion, here we setting $\mathcal{R} = 10$, $G=M_p=1$ as an example. Thus, one obtain that $M_c  = 2.66667$, $\beta _c  = 4.26667$ and $T_c  = 0.01827$. According to Eq.~(\ref{eq18}) and Eq.~(\ref{eq20}), the corresponding ``$T_{{\rm{local}}}^{{\rm{modif\/ied}}}  - M$'' diagram of rainbow SC black hole for dif\/ferent $\beta_0$ is depicted in Fig.~\ref{fig4}.

Fig.~\ref{fig4} depicts $T_{{\rm{local}}}^{{\rm{modif\/ied}}}-M$  as a function of mass for f\/ixed  $G = M_p  = 1$ and $\mathcal{R}=10$. The value of $\beta_0$ increases from top to bottom. The red solid curve, black dashed curve and blue dotted curve correspond to the critical GUP parameter  $\beta _0  = 0.234\beta _c  \approx 1$, $\beta _0  = \beta _c  = 4.26667$, and  $\beta _0  = 1.17\beta _c  \approx {\rm{5}}$, respectively. Obviously, one can observe a phase transition for $0 < \beta _0  < \beta _c$. Hence, in order to study the phase transition and critical phenomena of rainbow SC black hole, we set $\beta _0  = 1$ in the following discussions.

\begin{figure}[htbp]
\centering
\includegraphics[width=.5\textwidth,origin=c,angle=0]{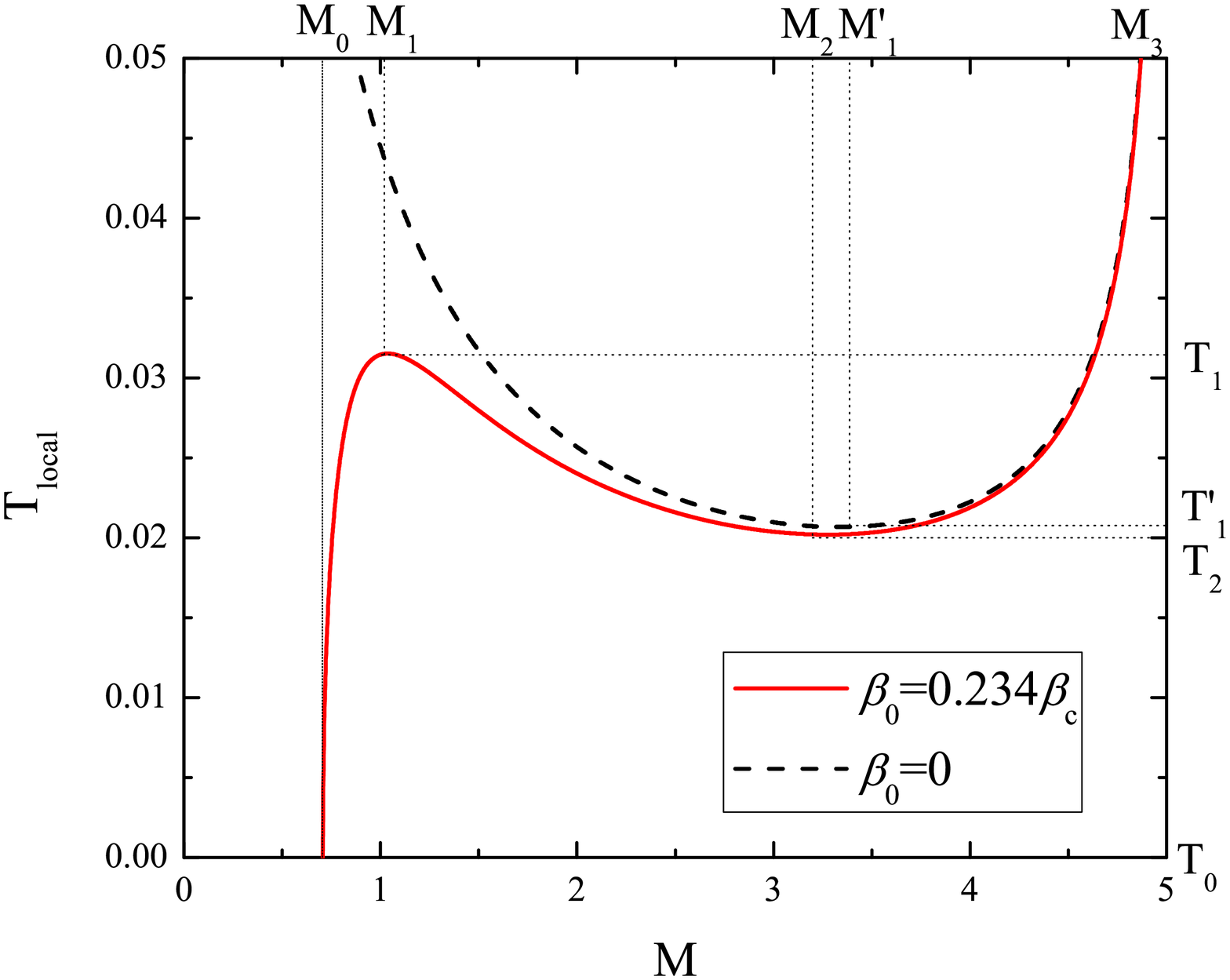}
\caption{\label{fig5} The original and modif\/ied local temperature versus mass at $G=M_p=1$ and $\mathcal{R}=10$. }
\end{figure}

\begin{figure}[H]
\centering
\includegraphics[width=.5\textwidth,origin=c,angle=0]{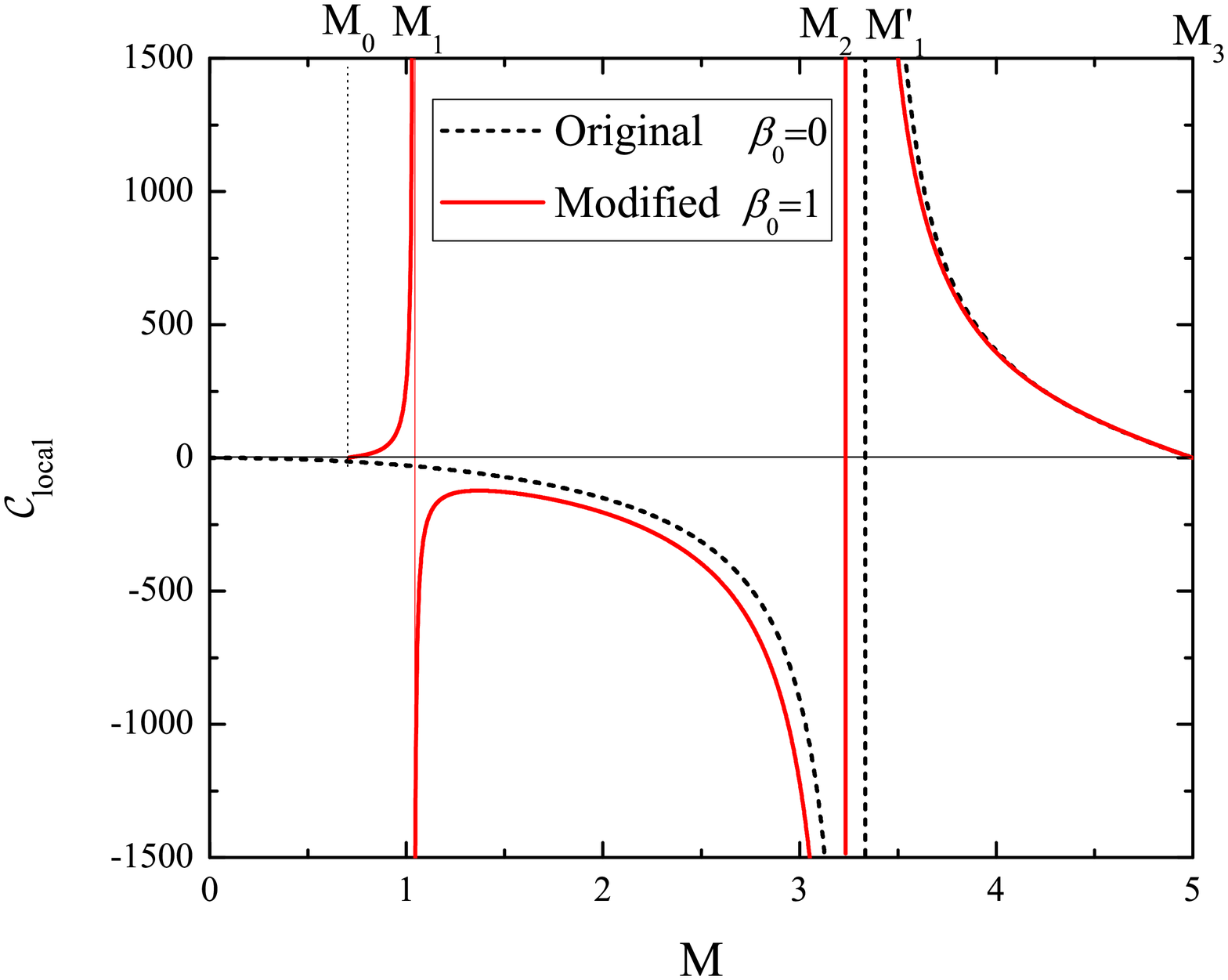}
\caption{\label{fig6}  The original and modif\/ied local heat capacity versus mass with $G= M_p=1$ and $\mathcal{R}=10$. }
\end{figure}

Next, let us plot the original and rainbow local temperature versus mass for $G = M_p =1$ and $\mathcal{R} = 10$ in Fig.~\ref{fig5}. The black dashed curve represents the local temperature of original SC black hole, and the red solid curve represents the local temperature of rainbow black hole. The evaporation process of SC black hole in the framework of rainbow function is commonly organized in three stages: at the f\/irst stage ($M_2 \leq M \leq M_3$), the rainbow local temperature decreases through its evaporation process; at the second stage ($M_1 \leq M \leq M_2$), the rainbow local temperature increases as mass decreases, lasting up to the mass in which it comes near to a maximum value; at the f\/inal stage ($M_0 \leq M \leq M_1$), the mass of rainbow SC black hole cannot get smaller than $M_0$ since the negative temperature becomes violates the laws of thermodynamics, which indicates that the RG can stop the black hole evolution and leads to a remnant. Hence, we have ${M_0} = {M_{{\rm{res}}}}$.
Employing the f\/irst law of thermodynamics, expression of total internal energy is given by
\begin{align}
\label{eq21}
E_{{\rm{local}}}^{{\rm{modif\/ied}}} & =  \int_{M_{0} }^M {T_{{\rm{local}}}^{{\rm{modif\/ied}}} {\rm{d}}S} 
 \nonumber \\
& = \frac{\mathcal{R}}{G}\left( {\sqrt {1 - \frac{{M_p G\sqrt {2\beta _0 } }}{\mathcal{R}}}  - \sqrt {1 - \frac{{2GM}}{\mathcal{R}}} } \right).
\end{align}
When $\beta _0  \to 0$, the local energy of original SC black hole $E_{\rm{local}}  = {{\mathcal{R}\left( {1 - \sqrt {1 - {{2GM} \mathord{\left/ {\vphantom {{2GM} r}} \right.  \kern-\nulldelimiterspace} \mathcal{R}}} } \right)} \mathord{\left/ {\vphantom {{\mathcal{R}\left( {1 - \sqrt {1 - {{2GM} \mathord{\left/ {\vphantom {{2GM} r}} \right.
 \kern-\nulldelimiterspace} \mathcal{R}}} } \right)} G}} \right. \kern-\nulldelimiterspace} G}$ is recovered. Then, using the modif\/ied local temperature Eq.~(\ref{eq18}) and the total internal energy Eq.~(\ref{eq21}), the modif\/ied heat capacity at f\/ixed ${\mathcal{R}}$ can be expressed as
\begin{align}
\label{eq22}
{\cal C}_{{\rm{local}}}^{{\rm{modif\/ied}}} &  = \left( {\frac{{\partial E_{{\rm{local}}}^{{\rm{modif\/ied}}} }}{{\partial T_{{\rm{local}}}^{{\rm{modif\/ied}}} }}} \right)_\mathcal{R} 
 \nonumber \\
& = \frac{{8\pi GM^3 \left( {\mathcal{R} - 2GM} \right)\sqrt {4M^2  - 2\beta _0 M_p^2 } }}{{M\left( {6GM^2  - 2M\mathcal{R} - 5\beta _0 GM_p^2 } \right) + 2\mathcal{R}M_p^2 \beta _0 }}.
\end{align}
Clearly, when $\beta _0$  vanishes, the Eq.~(\ref{eq22}) reduces to the local heat capacity of original SC black hole ${C_{{\rm{local}}}} = 8\pi G{M^2}{{\left( {\mathcal{R} - 2GM} \right)} \mathord{\left/ {\vphantom {{\left( {\mathcal{R} - 2GM} \right)} {\left( {3GM - \mathcal{R}} \right)}}} \right.  \kern-\nulldelimiterspace} {\left( {3GM - \mathcal{R}} \right)}}$ \cite{ch40}. According to Eq.~(\ref{eq22}), the variation of the heat capacity $\mathcal{C}_{{\rm{local}}}^{{\rm{modif\/ied}}}$ with the mass $M$ is plotted in Fig.~\ref{fig6}.
As shown in Fig.~\ref{fig6}, the rainbow heat capacity of SC black hole (red solid line) changes its sign at $M_1$ and $M_2$ for which the denominator in Eq.~(\ref{eq22}) vanishes. Meanwhile, the rainbow heat capacity $\mathcal{C}_{\rm rainbow}$ goes to zero when mass approaches $M_0$. When $M \rightarrow M_3$, the rainbow heat capacity tends to the original case (black dashed line) since the ef\/fect of GUP is negligible at this point. The rainbow heat capacity is divergent at the extreme points $M_1$ and $M_2$. Hence, the phase transitions near there are second-orders. It is well known that the positive heat capacity means the thermally stable state, while the negative heat capacity is considered to be the unstable solution. So, it is easy to f\/ind that the rainbow black hole has one unstable region (${M_1} \le M \le {M_2}$), and two stable regions (${M_0} \le M \le {M_1}$  and ${M_2} \le M \le {M_3}$). Now, the rainbow black hole can be classif\/ied into three types based on their local temperature and the heat capacity, namely, a large black hole (LBH) for ${M_2} \le M \le {M_3}$, a small black hole (SBH) for ${M_0} \le M \le {M_1}$, and a intermediate black hole (IBH) for ${M_1} \le M \le {M_2}$, which never appears in the original case. Obviously,  the small black hole (SBH) and the large black hole (LBH) are stable, showing that they can survive for a long time within the frame of the RG theory. On the contrary, the IBH is unstable because its heat capacity is negative, which implies that it would decay into the SBH or LBH quickly. In addition, it should be noted that the stability of the black holes can be known from their free energy.

\begin{figure}[htbp]
\centering
\includegraphics[width=.5\textwidth,origin=c,angle=0]{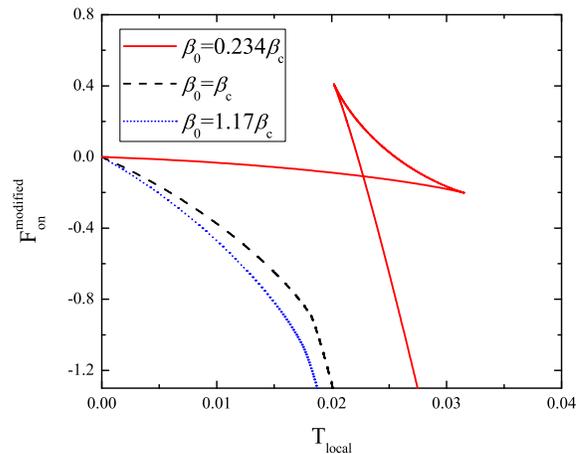}
\caption{\label{fig7} Variation in free energy of rainbow SC black hole with local temperature for dif\/ferent $\beta_0$ with $G= M_p=1$ and $\mathcal{R}=10$. }
\end{figure}

In order to obtain more details of the thermodynamic phase transition of the rainbow SC black hole, we use the three states of rainbow SC black hole to analyze the phase transition. In Refs.~\cite{ch41,ch42}, the free energy is defined as $F_{{\rm{on}}}  = E_{{\rm{local}}}  - T_{{\rm{local}}} S$. Therefore, the free energy of rainbow SC black hole is given by
\begin{align}
\label{eq22+}
 F_{{\rm{on}}}^{{\rm{modif\/ied}}} & = \frac{{\cal R}}{G}\left( {\sqrt {1 - \frac{{{M_p}G\sqrt {2{\beta _0}} }}{{\cal R}}}  - \sqrt {1 - \frac{{2GM}}{{\cal R}}} } \right)
  \nonumber \\
& - \frac{{\sqrt 2 M\left( {2{M^2} - M_p^2{\beta _0}} \right) + \Delta }}{{4{M^2}\sqrt {2 - {{4GM} \mathord{\left/ {\vphantom {{4GM} {\cal R}}} \right. \kern-\nulldelimiterspace} {\cal R}}} }}.
 \end{align}
where $\Delta  = M_p^2{\beta _0}\sqrt {2{M^2} - {\beta _0}M_p^2} {\rm{ln}}\left( {2M + \sqrt {4{M^2} - 2{\beta _0}M_p^2} } \right)$. Note that Eq.~(\ref{eq22+}) is reduced to the original free energy for $\beta _0  = 0$. Meanwhile, as seen from Fig.~\ref{fig7} that the thermodynamic phase transition of the rainbow SC black hole occurs for $0 < \beta _0  < \beta _c$.

From Fig.~\ref{fig7}, we show $F_{{\rm{on}}}$ curves with $T_{{\rm{local}}}$ for dif\/ferent $\beta _0$. The value of GUP parameter $\beta _0$ increases from top to bottom. It is clear that the swallow tail structure appears when the GUP parameter  $\beta _0$ is smaller than the critical value $\beta _c$, which indicates there is a two-phase coexistence state. The results are consistent with the prof\/ile of $T_{{\rm{local}}}^{\rm{modif\/ied}}-M$ in Fig.~\ref{eq3}. Meanwhile, Fig.~\ref{fig7} indicates a reverse van der Waals like behavior for $T>T_{\rm c}$, which is completely dif\/ferent from the van der Waals one of physical thermodynamic systems in Ref.~\cite{ch40a}.

\begin{figure}[htb]
\centering
\subfigure[$\beta_0=0$]{\includegraphics[height=6.2cm,width=7.2cm]{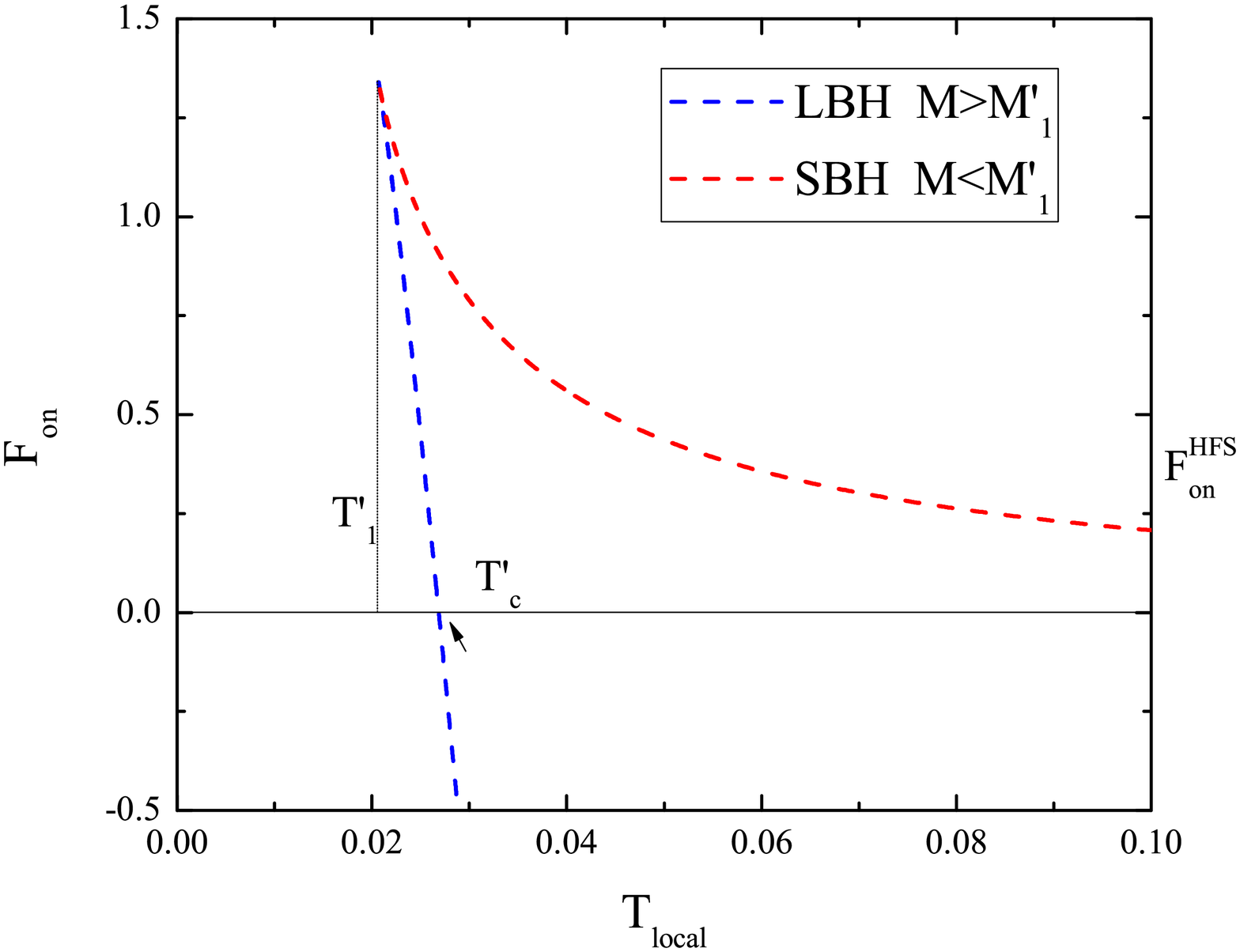}}
\subfigure[$\beta_0=1$]{\includegraphics[height=6.2cm,width=7.2cm]{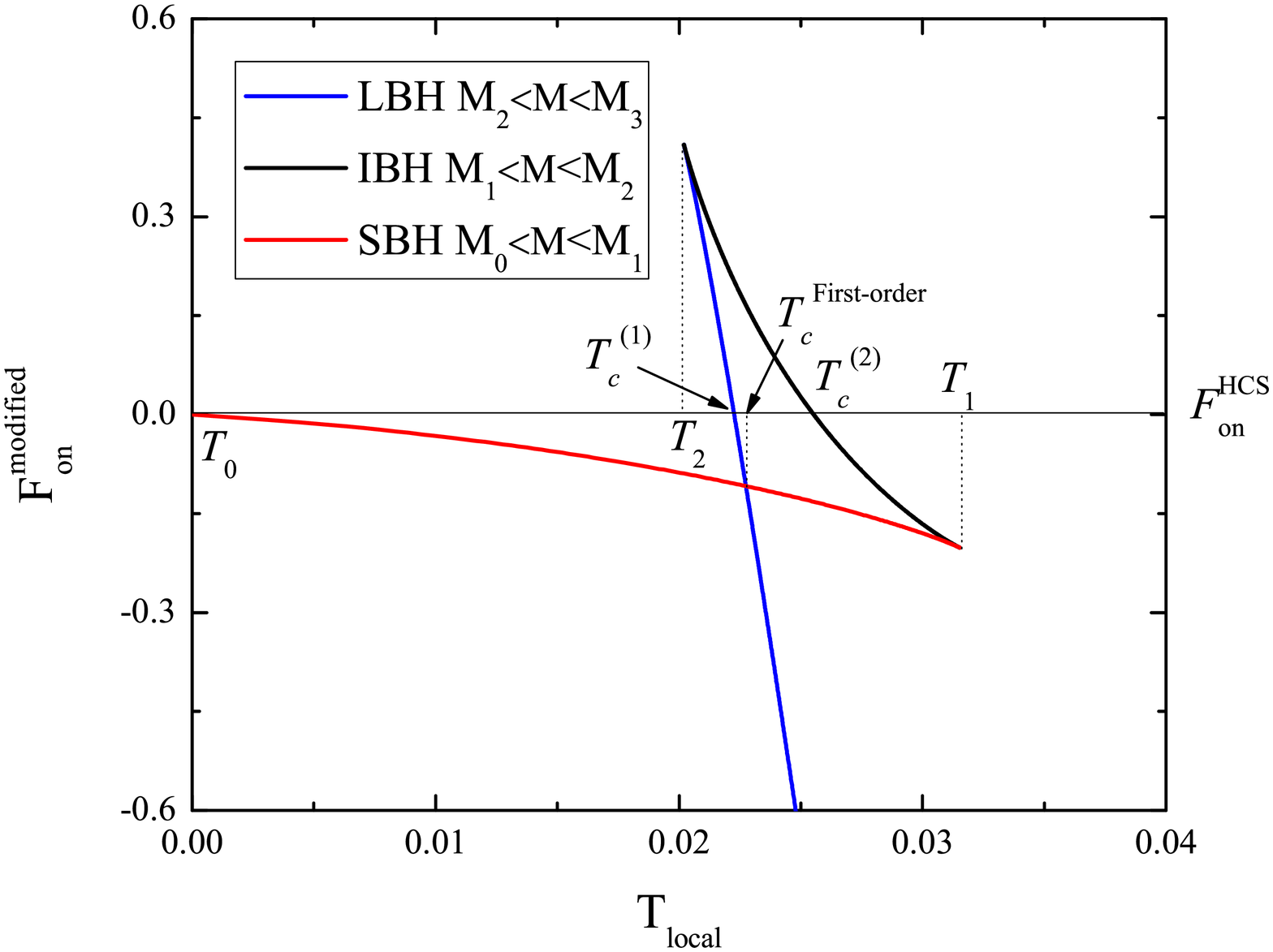}}
\caption{(a) shows the original free energy of SC black hole as function of the local temperatures.
(b) shows the modif\/ied free energy of SC black hole as function of the local temperatures. Here we choose $G = M_p = 1$ and $\mathcal{R}=10$.}
\label{fig8}
\end{figure}

For further investigate the phase transition between the black holes and the hot space, we plot Fig.~\ref{fig8}.

As seen from  Fig.~\ref{fig8}(a), it is found that the original free energy vanishes when $M \to {\rm{0}}$ because the vacuum state is Minkowski spacetime. Therefore, the $F_{\rm{on}}^{{\rm{HFS}}}$ represents the free energy of the hot f\/lat space (HFS). In addition, the local temperature has a minimal value $T_1'$, below which no black hole solution exists. The small-large black hole (Hawking-Page) transition occurs at $T_c'$. For  $T < T_c ^\prime$, both the free energy of small and large black holes are higher than the $F_{\rm{on}}^{{\rm{HFS}}}$, it means that the HFS is more probable than the SBH and LBH. However, for  $T > T_c ^\prime$, the free energy of SBH is higher than the $F_{\rm{on}}^{{\rm{HFS}}}$ while the LBH is lower than the $F_{\rm{on}}^{{\rm{HFS}}}$, which indicates that the LBH is more probable than the HFS. Therefore, one can find the radiation collapse to the LBH, and the SBH eventually decays into the LBH thermodynamically  \cite{cha1}.

Next, let us focus on Fig.~\ref{fig8}(b) for $G = \beta_0= M_p = 1$ and $\mathcal{R} = 10$. In contrast to the conventional one, the rainbow free energy becomes to zero when $M \rightarrow M_{\rm res} = M_0$, which implies the rainbow SC black hole only exits the hot curved space (HCS) \cite{ch43}. Therefore, we study phase transition between various black hole states and the HCS via analyzing the free energies of the rainbow SC black hole.

(i) The SBH and IBH are degenerate at $T_1$ with the mass $M_1$; the intermediate-large black hole transition occurs at the inf\/lection point $T_2$ corresponding to $M_2$. Interestingly, it is found two intersections points (Hawking-Page-type critical points) between the line of free energy and the $F^{\rm HFS}$ at $T^{(1)}_c$ and $T^{(2)}_c$ in Fig.~\ref{fig8}(b), while only one Hawking-Page phase transition in the original case, as seen from Fig.~\ref{fig8}(a).

(ii) There is a one-order phase transition at critical temperature $T^{\rm f\/irst-order}_c$ since the images shows a characteristic swallow tail behavior.

(iii) For  $T_0  < T < T_c^{(1)}$, the free energy of IBH and LBH are higher than the  $F_{\rm{on}}^{{\rm{HCS}}}$, whereas the SBH is lower than $F_{\rm{on}}^{{\rm{HCS}}}$, it implies that the stable SBH is more probable than HFS. For $T_c^{(1)}  < T < T_c^{\rm f\/irst-order}$, both SBH and LBH are lower than the $F_{\rm{on}}^{{\rm{HCS}}}$. It should be noted that LBH and IBH should decay into SBH since the free energy of SBH is lower than those of IBH and LBH in this region. Consequently, for $T_0 < T < T_c^{\rm f\/irst-order}$, SBH should undergoes a tunneling, which indicates that the HFS can collapse to SBH. As the local temperature increase, one f\/inds that free energies of black holes satisfy the relation $F_{\rm{on}}^{{\rm{LBH}}} < F_{\rm{on}}^{{\rm{SBH}}} < F_{\rm{on}}^{{\rm{HCS}}} < F_{\rm{on}}^{{\rm{IBH}}}$ for  $T_c^{\rm f\/irst-order} < T < T_c^{\left( 2 \right)}$. Then, the relation becomes $F_{\rm{on}}^{{\rm{LBH}}} < F_{\rm{on}}^{{\rm{SBH}}} < F_{\rm{on}}^{{\rm{IBH}}} < F_{\rm{on}}^{{\rm{HCS}}} $ for  $T_c^{\left( 2 \right)}  < T < T_1$, it means that HCS does not only collapse into SBH but also into IBH and LBH. However, the IBH with negative heat capacity is unstable. Therefore, it would decay into the SBH. Meanwhile, it is obvious that the free energy of SBH is always higher than free energy of the LBH, it leads to the SBH eventually decays into the LBH thermodynamically. Therefore, for  $T_c^{\rm f\/irst-order}  < T < T_1$, the HCS would eventually decays into the LBH via the quantum tunneling ef\/fect.

\section{Discussion and conclusion}
\label{Dis}
In this paper, we f\/irst derived a new kind of rainbow functions, which contains the GUP parameter. Then, we considered this rainbow functions to investigate the modif\/ied thermodynamics and phase transitions the rainbow SC black hole. F\/irst of all, our results showed that the rainbow functions change the picture of Hawking radiation and stop the evaporation of a black hole as its size approaches the Planck scale. It naturally leads to a remnant of SC black holes, which can be considered as a candidate of dark matter. Second, as seen from Fig.~\ref{fig4}, the phase transition occurs when the GUP parameter is smaller than the critical points. Therefore, we have set the rainbow parameters as $\beta _0  = 1$ in order for the study to be focused and feasible. Third, since the heat capacity enjoys two divergencies in Fig.~\ref{fig6}, it demonstrated the thermodynamic system of rainbow SC black hole has two second order phase transitions at   $M_1$ and $M_2$. Fourth, according to the ``$F_{{\rm{on}}}  - T_{{\rm{local}}}$'' diagram, one can find an unstable black hole (IBH) interpolating between the stable SBH and LBH, which never exists in the original case. Fifth, the $F_{{\rm{on}}}^{{\rm{modif\/ied}}}$ surface showed the swallow tail shape, which indicates that the thermodynamic system of rainbow SC black hole exists a f\/irst-order transition. Sixth, it is found two Hawking-Page-type critical points from Fig.~\ref{fig8}(b), while only one Hawking-Page critical point in the original case. Finally, for $T_0  < T < T_c^{\rm {First-order}}$, since the free energy of SBH is lower than those of the IBH and LBH, the unstable IBH and stable LBH decay into the stable SBH. However, for $T_c^{\rm {First-order}} < T < T_1$, it is obviously the free energy of stable LBH is always lower than those of the SBH and IBH, this leads to the SBH and IBH eventually collapse into the LBH thermodynamically.

\vspace*{3.0ex}
{\bf Acknowledgements}
\vspace*{1.0ex}

This work is supported in part by the National Natural Science Foundation of China (Grant Nos. 11847048 and 11573022) and the Fundamental Research Funds of China West Normal University (Grant Nos. 17E093, 17YC518 and 18Q067).

\end{document}